\documentclass[aps,prl,twocolumn]{revtex4}

\usepackage{graphicx}
\usepackage{latexsym}
\usepackage{amsmath}
\usepackage{amssymb}
\usepackage{psfrag}

\begin{document}

\title{A simple and efficient approach to the optimization of correlated wave functions}
\author{Anthony Scemama and Claudia Filippi}
\affiliation{
Instituut-Lorentz, Universiteit Leiden, Niels Bohrweg 2, 2333 CA Leiden,
The Netherlands}
\date{\today}

\begin{abstract}
We present a simple and efficient method to optimize within 
energy minimization the determinantal component of the many-body
wave functions commonly used in quantum Monte Carlo calculations. 
The approach obtains the optimal wave function as an 
approximate perturbative solution of an effective Hamiltonian 
iteratively constructed via Monte Carlo sampling.
The effectiveness of the method as well as its ability to substantially
improve the accuracy of quantum Monte Carlo calculations is demonstrated 
by optimizing a large number of parameters for the ground state of acetone 
and the difficult case of the $1{}^1\mbox{B}_{1u}$ state of hexatriene.
\end{abstract}

\maketitle

Over the last decade, quantum Monte Carlo (QMC) methods have been 
employed to accurately compute electronic properties of large molecular 
and solid systems where conventional quantum chemistry approaches are 
extremely difficult to apply~\cite{review}. 
A crucial step in both the variational (VMC) and the diffusion Monte 
Carlo (DMC) approach is the construction of the trial wave function
$\Psi_{\rm T}$ which is usually chosen of the Jastrow-Slater form, that is
$\Psi_{\rm T}={\cal J}\Phi$, where $\Phi$ is a small expansion in Slater 
determinants and ${\cal J}$ the positive Jastrow correlation factor. 

Although considerable progress has been made (principally using the 
variance minimization approach~\cite{varmin}) in the construction 
of optimal Jastrow factors, relatively little attention has been given 
to the optimization of the determinantal part of the wave
function. Methods such as Hartree-Fock (HF) or a small scale configuration 
interaction (CI) are used as a practical 
way of constructing the determinantal component, which
is generally not reoptimized when the Jastrow factor is added.
However, the determinantal part of the wave function solely determines 
the DMC energy~\cite{non_loc_pseudo} and often needs to be 
reoptimized to obtain accurate results~\cite{SchautzFilippi04}.
A practical and simple approach to calculate the optimal
determinantal component is therefore particularly important 
to a wide and successful application of QMC methods.

In recent years, several methods to optimize the wave function through 
energy minimization have been proposed~\cite{Harju97,
Snajdr99,Fahy99,FilippiFahy00,Rappe,Nightingale01,Sorella01,
SchautzFahy02,PrendergastFahy02,CasulaSorella03, 
SchautzFilippi04,UmrigarFilippi05,Sorella05}.
A direct approach to energy minimization
entails to compute the gradient and the Hessian of the energy 
with respect to the desired parameters.
The use of an estimate of the Hessian characterized 
by reduced statistical fluctuations~\cite{UmrigarFilippi05,Sorella05}
yields a simple and robust optimization algorithm for the 
Jastrow parameters.
However, the Hessian with respect to the orbital parameters in the
determinant is affected by higher statistical noise
so that devising a stable energy-minimization scheme 
is more difficult~\cite{Sorella05}: the resulting approach 
is for instance less stable that the simple stochastic reconfiguration (SR) 
method~\cite{Sorella01,CasulaSorella03} and, during the optimization, 
may have to reduce to the inefficient SR to retain stability.
To date, the most successful method 
remains the energy fluctuation potential (EFP) 
method~\cite{Fahy99,FilippiFahy00,SchautzFahy02,PrendergastFahy02,
SchautzFilippi04} which determines the
optimal determinantal component as the solution of an effective Hamiltonian
iteratively constructed via Monte Carlo sampling. 
The approach has been used to optimize the 
orbitals~\cite{FilippiFahy00,SchautzFilippi04}, the linear coefficients 
in front of the determinants~\cite{SchautzFahy02,SchautzFilippi04},
and has been extended to excited states~\cite{SchautzFilippi04}. 
The method is very stable and more efficient
than the SR approach but quite complex~\cite{SchautzFilippi04}:
the construction of the starting effective Hamiltonian as well as its update 
are computationally very demanding steps, especially for large systems.

In this Letter, we propose a simple and efficient optimization method
for the determinantal component of the wave function. The approach 
constructs the optimal wave function via a perturbative scheme based 
on the EFP method, and only requires easily accessible quantities from 
the quantum chemical calculation used to set up the starting wave 
function.  The performance of the method is demonstrated on
acetone and on the difficult case of the $1{}^1\mbox{B}_{1u}$ 
state of hexatriene.
First, we will briefly review the EFP method.

{\it The energy fluctuation potential method}.
Let us assume that the trial wave function $\Psi_{\rm T}$ depends on a set 
of parameters $\{\alpha_k\}$. The derivatives of the energy with respect to
the parameters can be written as:
\begin{eqnarray}
\frac{\partial E}{\partial \alpha_k} =
2 \langle (E_{\rm L} - \bar{E}) (O_k - \bar{O}_k) \rangle \,,
\label{eq:correlator}
\end{eqnarray}
where $\langle \cdot \rangle$ denotes the average with respect to the
square of the trial wave function $|\Psi_{\rm T}|^2$, which can be computed 
by Monte Carlo sampling. We defined $\bar{E} = \langle E_{\rm L} \rangle$ and 
$\bar{O}_k = \langle O_k \rangle$ where
\begin{equation}
E_{\rm L} = \frac{{\cal H} \Psi_{\rm T}}{\Psi_{\rm T}} \qquad {\rm and} \qquad 
O_k = \frac{1}{ \Psi_{\rm T}}\frac{\partial \Psi_{\rm T}}{\partial \alpha_k}
\,.
\label{eq:defO}
\end{equation}
For the optimal parameters, the derivatives of the energy 
(Eq.~\ref{eq:correlator}) are zero, and the fluctuations of the local energy 
$E_{\rm L}$ and of the functions $O_k$ are uncorrelated.
This means that the energy is stationary if the remaining fluctuations of 
the local energy cannot be further reduced by adding some combination of 
the functions $O_k$.
Hence, the minimization of the energy can be reformulated as a least-squares
fit of the fluctuations of the local energy with the 
functions $O_k$:
\begin{equation}
\chi^2 = \langle (E_{\rm L} - V_0 - \sum_{k>0} V_k O_k)^2 \rangle\,,
\label{eq:chi2}
\end{equation}
A minimization of $\chi^2$ with respect to the parameters $V_k$ leads to the
following set of linear equations for $m>0$:
\begin{equation}
\langle \Delta E \Delta O_m \rangle=
\sum_{k>0} V_k \langle \Delta O_k \Delta O_m \rangle
\label{eq:delta}
\end{equation}
where we have eliminated $V_0$ from the other equations, and 
$\Delta E = E_{\rm L} - \bar{E}$ and $\Delta O_m = O_m - \bar{O}_m$. 
Since the left-hand side of these equations corresponds to the derivatives 
of the energy (Eq.~\ref{eq:correlator}), the 
fitting parameters $V_k$ are zero if and only if all the derivatives of the 
energy are zero.
In general, the parameters $V_k$ which solve these linear equations 
will not be equal to zero and can be used to improve the current trial
wave function $\Psi_{\rm T}$. 

We focus here on the EFP procedure to optimize the determinantal 
part $\Phi$ of the trial wave function $\Psi_{\rm T}$:
\begin{equation}
\Psi_{\rm T} = {\cal J} \Phi = {\cal J} \sum_i c_i C_i \,,
\label{eq:csf_j}
\end{equation}
where a spin-adapted configuration state function (CSF) $C_i$ is a linear 
combination of Slater determinants $D_i$ of single-particle orbitals.
Let us assume that the starting $\Phi$ is the lowest solution 
$\Phi_0^{(0)}$ of the CI Hamiltonian ${\cal H}^{(0)}$:
\begin{equation}
{\cal H}^{(0)}=\sum_i E_i^{(0)} | \Phi_i^{(0)}\rangle\langle \Phi_i^{(0)} |\,.
\end{equation}
where the states $\Phi_i^{(0)}$ span the same space as the CSFs.
To obtain the optimal coefficients $c_i$ or equivalently the 
best solution in the basis of the eigenstates of ${\cal H}^{(0)}$, 
we consider the variations of $\Phi$ with respect to the 
eigenstates other than $\Phi^{(0)}_0$: 
\begin{eqnarray}
\Phi=\Phi_0^{(0)} \rightarrow \Phi'=\Phi_0^{(0)}+
\sum_{k>0} \delta_k \Phi_k^{(0)}\,,
\label{eq:wfchange}
\end{eqnarray}
so that $O_k=\Phi_k^{(0)}/\Phi_0^{(0)}$.
The quantities 
appearing in the linear equations (Eq.~\ref{eq:wfchange}) are sampled 
from $\Psi_{\rm T}={\cal J}\Phi_{\rm 0}^{(0)}$ and parameters $V_k^{(0)}$ 
are used to construct a new Hamiltonian ${\cal H}^{(1)}$ as
\begin{equation}
{\cal H}^{(1)} = {\cal H}^{(0)} + \sum_{k > 0} V_k^{(0)}
\left( |\Phi_0^{(0)} \rangle \langle \Phi_k^{(0)}| + h.c. \right)\,.
\label{eq:hiter}
\end{equation}
This Hamiltonian is diagonalized, yielding a new set of states,
and the procedure is iterated until convergence.

If we also want to optimize the orbitals in the Slater determinants, 
we can linearize the problem using a so-called super-CI expansion
to treat the CI and orbital variations on the same footing:
the CSFs occupied in the wave function (Eq.~\ref{eq:csf_j}) are augmented 
by all possible single excitations from the occupied to a set of external 
orbitals, and
the occupied orbitals can be improved by using the natural 
orbitals of the CI wave function in this augmented space~\cite{Ruedenberg}.
The EFP scheme to optimize the determinantal component via a
super-CI approach~\cite{SchautzFilippi04} is quite complex.
Setting up the starting super-CI Hamiltonian 
is not trivial, and
the number of singly-excited CSFs increases with 
the number of CSFs in the reference wave function and with the size of the
basis (i.e.\ with the number of available external orbitals) so that the 
approach quickly becomes computationally too demanding. 

{\it A simplified EFP approach}. The basic idea of the scheme proposed
here is to avoid the explicit construction of the EFP Hamiltonian by solving 
the problem perturbatively. For the example given above, we obtain an improved 
wave function (Eq.~\ref{eq:wfchange}) not by diagonalizing the Hamiltonian 
(Eq.~\ref{eq:hiter}) but approximately, to first order in the perturbation 
given by the corrections $V_k^{(0)}$:
\begin{equation}
\Phi=\Phi_0^{(0)} \rightarrow \Phi'=\Phi_0^{(0)}-
\sum_{k>0} \frac{V_k^{(0)}}{E_k^{(0)}-E_0^{(0)}} \Phi_k^{(0)}\,.
\label{eq:pert}
\end{equation}
It is simple to show that, in the absence of the Jastrow factor,
this is indeed the perturbative solution for ${\cal H} -{\cal H}^{(0)}$ since
$V_k^{(0)}=\langle \Phi_k^{(0)}|{\cal H} -{\cal H}^{(0)} | \Phi_0^{(0)}
\rangle$.
This perturbative EFP scheme can also be viewed
as a generalization of the SR method~\cite{Sorella01,CasulaSorella03}
which yields an improved wave function by applying the 
operator $\Lambda-{\cal H}$ to the current state and projecting
the result onto the space defined by the parameterization. 
The SR wave function is constructed as in Eq.~(\ref{eq:pert}) but all
the changes are scaled by the same energy denominator
while the EFP approach scales each 
correction appropriately, achieving 
significantly faster convergence~\cite{SchautzFilippi04}. 

{\it Estimate of the energy denominators}.
The perturbative solution of the EFP problem
requires the knowledge of the energy denominators
appearing in Eq.~(\ref{eq:pert}) which can be easily estimated 
with the use of quantities readily available from the quantum chemical 
calculation used to set up the starting wave function.
We will first consider the optimization of the orbital parameters.

The simplest case is given by a trial wave function constructed from a 
single closed-shell determinant as $\Psi_{\rm T}={\cal J} D_0$, where
$D_0$ is a determinant of Hartree-Fock or density functional
Kohn-Sham orbitals which we want to reoptimize in the presence of the 
Jastrow factor. We assume here that we have chosen HF orbitals but 
all the following considerations equally apply to the case of density 
functional orbitals.
The set of occupied and virtual HF orbitals and the atomic 
basis on which they are expanded span the same space so that we can 
express the variations of one orbital with respect to the expansion 
coefficients in terms of the variations with respect to the other 
orbitals. To optimize the $M$ occupied orbitals $\varphi_i$ 
in $D_0$, we only need to consider the variations with respect to the 
$N-M$ virtual orbitals as
\begin{eqnarray}
\varphi_i \rightarrow \varphi_i'=
\varphi_i + \sum_{j=M+1}^N c_{ij} \varphi_j\,,
\label{eq:mix_orb}
\end{eqnarray}
where we only mix orbitals of the same symmetry.
The corresponding first-order change in the wave function is 
\begin{eqnarray}
\Psi_{\rm T} \rightarrow 
\Psi_{\rm T}' &=& {\cal J}D_0 + \sum_{i=1}^{M} \sum_{j=M+1}^{N}
{c_{ij}} {\cal J} C_0^{i\rightarrow j} \,,
\end{eqnarray}
where $C_0^{i\rightarrow j}=D_0^{i\alpha\rightarrow j\alpha}+
D_0^{i\beta\rightarrow j\beta}$ is the CSF of the two 
determinants obtained by substituting the orbital $\varphi_i$ 
with orbital $\varphi_j$ for the up and down spin, respectively.
It is simple to show that 
$O^{i\rightarrow j} = C_0^{i\rightarrow j}/D_0$.

Following the EFP procedure, we sample from $\Psi_{\rm T}$ the 
quantities appearing in the linear equations (Eq.~\ref{eq:delta}) 
and obtain the corrections $V^{i\rightarrow j}$ corresponding to 
the functions $O^{i\rightarrow j}$.  We can then obtain an 
improved determinantal component in analogy to 
Eq.~(\ref{eq:pert}) as
\begin{equation}
\Phi' = D_0 - \sum_{i=1}^{M}  \sum_{j=M+1}^{N}
\frac{V^{i\rightarrow j}}{\Delta E^{i\rightarrow j}} 
C_0^{i\rightarrow j}\,, 
\label{eq:pert_1det}
\end{equation}
where we assign an energy scale $\Delta E^{i\rightarrow j}$ 
to the variation corresponding to the single excitation 
$C_0^{i\rightarrow j}$. We now make the key observation that,
to first order, this is equivalent to the much simpler step of
constructing an improved set of occupied single-particle orbitals 
as
\begin{eqnarray}
\varphi_i'=\varphi_i - \sum_{j=M+1}^N \frac{V^{i\rightarrow j}}
{\Delta E^{i\rightarrow j}}\varphi_j\,,
\label{eq:pert_1det_orb}
\end{eqnarray}
which can be used in a new single-determinant wave 
function $\Psi_{\rm T}' = {\cal J} D_0'$. We then proceed 
iteratively by sampling a new set of $V^{i\rightarrow j}$ from the wave 
function $\Psi_{\rm T}'$ and updating the orbitals as in 
Eq.~(\ref{eq:pert_1det_orb}) until convergence. 

To estimate the energy denominators
$\Delta E^{i\rightarrow j}$, we note that the starting HF
orbitals are canonical orbitals obtained by diagonalizing the 
Fock matrix so that the corresponding eigenvalues have a 
physical interpretation via the Koopman's theorem.
Minus the eigenvalues of the occupied and the virtual orbitals are
an approximation to the ionization potentials (IP) 
and the electron affinities (EA), respectively.
The energy associated to the promotion of one electron from orbital 
$\varphi_i$ to orbital $\varphi_j$ can therefore be estimated
as the difference IP$-$EA, namely
\begin{equation}
\Delta E^{i\rightarrow j}
\sim E^{i\rightarrow j} - E_{\rm HF} \sim \epsilon_j - \epsilon_i\,,
\label{eq:deltaE_hf}
\end{equation}
where $\epsilon_i$ and $\epsilon_j$ are the eigenvalues corresponding
to orbitals $\varphi_i$ and $\varphi_j$.
We do not find it necessary at successive iterations to update the 
parameters $\Delta E^{i\rightarrow j}$ which are therefore
kept fixed at the starting HF values.

\begin{figure}[t]
\includegraphics[width=\columnwidth]{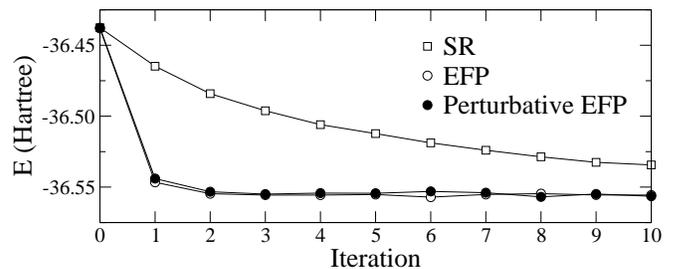}
\caption{Convergence of the VMC energy of acetone with the 
perturbative EFP, the EFP and the SR method to optimize the 473 orbital 
parameters. The starting determinantal component is from an unconverged HF 
calculation. The statistical errors are smaller than the symbol size.}
\label{fig:orb_1det}
\end{figure}

In Fig.~\ref{fig:orb_1det}, we compare the convergence of the proposed 
method, denoted by ``perturbative EFP'', with the EFP and the SR methods 
for the orbital optimization of a single closed-shell Jastrow-Slater 
wave function for the ground state of acetone (C$_3$H$_6$O).
To demonstrate the robustness of the method, we construct a poor 
starting determinantal component from a set of unconverged HF orbitals 
obtained by diagonalizing the Fock matrix built from the Huckel 
orbitals. A simple electron-electron and electron-nucleus Jastrow
factor is employed and the 473 orbital parameters are optimized using 
the starting unconverged eigenvalues as estimates for the energy 
denominators (Eq.~\ref{eq:deltaE_hf}). 
The SR optimization is performed with the critical value of 
$\Lambda$ estimated as in Ref.~\onlinecite{SchautzFilippi04}, and converges 
to the optimal wave function only after 40 steps. 
The perturbative EFP approach converges in only 2 steps 
as the EFP method, with a computational 
effort as low as the one of the SR method. Interestingly, 
the perturbative method is successful even when using the 
eigenvalues of the starting Huckel orbitals to roughly estimate 
the energies $\Delta E^{i\rightarrow j}$.
Finally, if we start from converged HF orbitals, the energy gain 
from the optimization is of 14 $\pm$ 1 and 6 $\pm$ 1 mHartree in VMC and
DMC, respectively.

The procedure is not very different if the wave function is 
constructed from a spin-unrestricted determinant. The variations of 
an occupied orbital (Eq.~\ref{eq:mix_orb}) are now with respect to 
the virtual orbitals of the same spin, and the excitation energies 
are estimated as
$\Delta E^{i,\sigma\rightarrow j,\sigma} \sim \epsilon_{j,\sigma} 
- \epsilon_{i,\sigma}$
where $\sigma$ is the spin of the varied orbital. 
The case of a restricted open-shell determinant is instead
more complicated since the final orbitals and eigenvalues of a restricted 
open-shell Hartree-Fock calculation are not uniquely defined.
Among the many constructions available in the literature,
we find that the one by Guest and Sauders~\cite{Guest74} yields 
eigenvalues which, in average, well approximate 
the excitation energies in Eq.~(\ref{eq:deltaE_hf}).
Note that closed-to-open-shell excitations must also
be included in Eq.~(\ref{eq:mix_orb}) to allow full orbital 
variation.

For a multideterminant wave function from a 
complete-active-space self-consistent-field (CASSCF) calculation, a 
set of canonical orbitals is usually defined as diagonalizing the
closed, active and virtual blocks of the so-called generalized 
Fock matrix.  However, to obtain a good estimate of the
closed-shell-to-active and active-to-virtual excitation energies, 
the eigenvalues must be properly adjusted since the energy of an 
active orbital should be different for excitations into or out of 
it, i.e.\ closer to an electron affinity or to an ionization 
potential, respectively.  Therefore, for these excitations, we 
follow Ref.~\onlinecite{Ghigo04} to define 
$\Delta E^{i\rightarrow j}\sim  E^{i\rightarrow j}-E_{\rm CASSCF}$
as:
\begin{eqnarray} 
\Delta E^{i\rightarrow j} 
&\sim & \epsilon_j - \epsilon_i +\frac{\lambda}{2}\left( \rho_{jj} 
+ 2 - \rho_{ii} \right)\label{eq:deltaE_mcscf}
\label{eq:de_cas}
\end{eqnarray} 
where $\epsilon$ is a CASSCF eigenvalue, $\rho$ is the single-particle 
density matrix, and $\lambda$ is an average difference between the 
electron affinity and the ionization potential of the active orbitals, 
chosen of the order of 0.3-0.5 Hartree.

Finally, we consider the simultaneous optimization of orbitals 
and CI coefficients.
The simplest approach is to alternate between an orbital 
optimization step with the perturbative EFP method, and a CI 
calculation  in the basis of the CSFs multiplied by the Jastrow 
factor~\cite{Nightingale01}.
Alternatively, we can use the perturbative EFP method for 
orbital and CI parameters and sample the quantities in 
Eq.~(\ref{eq:delta}) for both variations. The corrections 
$V_k$ for the CI coefficients are computed in the basis of the 
CI states obtained in a starting CI calculation, and the
energy denominators (Eq.~\ref{eq:pert}) 
estimated using the CI energies.

\begin{table}
\begin{tabular}{llccc}
\hline
State & Wave function & E$_{\rm VMC}$  & E$_{\rm DMC}$  & $\Delta$E (eV) \\
\hline
1$^1$A$_g$       
 & HF            & -38.684(1)     & -38.7979(7)    & -- \\
 & B3LYP         & -38.691(1)     & -38.7997(7)    & -- \\
 & optimized     & -38.691(1)     & -38.7992(7)    & -- \\
1$^1$B$_{1u}$        
 & CAS(2,2)      & -38.472(1)     & -38.5910(7)    & 5.63(3)   \\
 & B3LYP         & -38.482(1)     & -38.6030(7)    & 5.35(3) \\
 & optimized     & -38.493(1)     & -38.6069(8)     & 5.23(3) \\
expt.~\cite{Leopold} &  &  &                       & 5.22\hspace*{1.2em} \\
\hline
\end{tabular}
\caption{VMC and DMC energies in Hartree for the 1$^1$A$_g$ and
the 1$^1$B$_{1u}$ states of trans-hexatriene. The DMC excitation energies
$\Delta$E are with respect to the ground state obtained with the same 
wave function type. We optimize 604 and 493 parameters for the ground 
and excited state, respectively. The errors on the last figure are
given in parenthesis.}
\label{tab:hexatriene}
\end{table}

In Table~\ref{tab:hexatriene}, we demonstrate the performance of 
the method on the 1$^1$B$_{1u}$ state of trans-hexatriene (C$_6$H$_8$) 
which represents a challenge for all electronic structure approaches: 
CASPT2 yields a vertical excitation 
energy of only 5.01 eV~\cite{Serrano} while the TDDFT energies range 
between 4.42 and 4.64 eV, depending on the functional employed~\cite{Hsu}. 
For the single-determinant closed-shell ground state, the VMC energy is
improved by about 7 mHartree when otimizing the HF orbitals while,
differently from acetone, the DMC energy is not significantly 
affected~\cite{hexatriene}. The excited state is mainly an excitation from the highest 
occupied (LUMO) to the lowest unoccupied (HOMO) orbital and well described 
a two-determinant singlet wave function. A CASSCF wave function of two 
electrons in two orbitals, denoted by CAS(2,2), gives a DMC vertical 
excitation energy 0.4 eV higher than the experimental value. Employing
B3LYP orbitals partially improves the result but only the fully 
optimized wave function yields an excitation energy in perfect 
agreement with experiments. 
Finally, we note that both ground and excited states do not have  
multiconfigurational character and using an unoptimized multideterminant 
wave function yields significantly worse QMC energies: for the 
excited state, a CASSCF wave function of 6 electrons in 6 orbitals,
where we keep the leading 5 CSFs,
yields an energy of -38.438(1) and -38.560(2) Hartree in VMC and DMC, 
respectively. Upon optimization of both orbitals and CI coefficients, 
we recover the best energies of Table~\ref{tab:hexatriene}.

We thank C. J. Umrigar for useful discussions.
This work is 
sponsored by the Stichting Nationale Computerfaciliteiten
(NCF) for the use of supercomputer facilities.


\begin{thebibliography}{999}
\bibitem{non_loc_pseudo} With non-local pseudopotentials, 
the full Jastrow-Slater wave function is used to localize 
the pseudopotential so that the DMC energy will depend on the Jastrow factor.
\bibitem{hexatriene} An electron-electron and electron-nucleus Jastrow factor
is employed. The Gaussian basis set is (10$s$10$p$2$d$)/[4$s$4$p$2$d$] for
C and (11$s$2$p$)/[4$s$2$p$] for H. The DMC time-step is 0.025 Hartree$^{-1}$.
\end{thebibliography}
\end{document}